 \documentclass[final,3p,times,authoryear]{elsarticle}

\usepackage{amssymb}
\usepackage{amsmath}
\usepackage{booktabs}
\usepackage{array}
\usepackage{geometry}
\usepackage{graphicx}
\usepackage{pdflscape}
\usepackage{lscape,graphicx}
\usepackage[absolute]{textpos}
\usepackage{fancyhdr}
\usepackage{subcaption}
\usepackage{setspace}
\usepackage[colorlinks,citecolor=blue,linkcolor=blue,pdftex,breaklinks=true]{hyperref}
\setcitestyle{square,comma,sort&compress,numbers}
\usepackage{multirow}
\usepackage{appendix}

\journal{Nuclear Physics A}

\biboptions{sort&compress}

\begin{document}

\begin{frontmatter}



\title{Optical potential parameters of light nuclear fusion based on precise Coulomb wave functions}


\author{Binbing Wu$^{\rm a}$}
\author{Hao Duan$^{\rm b,c}$}
\author{Jie Liu$^{\rm a,d}$\footnote{Corresponding author: jliu@gscaep.ac.cn}}

\address{$^{\rm a}$Graduate School, China Academy of Engineering Physics, Beijing 100193, China}
\address{$^{\rm b}$Institute of Applied Physics and Computational Mathematics, Beijing 100088, China}
\address{$^{\rm c}$Laboratory of Computational Physics, Institute of Applied Physics and Computational Mathematics, Beijing 100088, China}
\address{$^{\rm d}$CAPT, HEDPS, and IFSA Collaborative Innovation Center of MoE, Peking University, Beijing 100871, China}
\begin{abstract}
Based on precise Coulomb wave functions (CWFs), we attempt to  calculate the fusion cross sections of light nuclei in a complex spherical square-well nuclear potential (i.e., optical potential model). Comparing with experimental benchmark cross section data, we can  calibrate  optical potential parameters associated with D+D, D+T, D+$^{3}{\rm He}$, p+D, p+$^{6}{\rm Li}$ and p+$^{7}{\rm Li}$ fusion reactions.
Surprisingly, we find that our calculated optical potential parameters are quite
different from  those of many previous results (e.g., Phys.\ Rev.\ C.\ {\bf 61} (2000) 024610,  Nucl.\ Phys. A {\bf 986} (2019) 98, etc. ), in which approximate Coulomb wave functions (ACWFs)  with only retaining the leading terms are exploited for the continuity conditions at the radius of nuclear potential.
Furthermore, with the obtained  optical potential parameters, we compare the fusion cross sections and astrophysical~$S$-factors with that formulated from  ACWFs approach, and also find  apparent deviations especially for the fusion reactions with resonance peaks such as  D+T and D+$^{3}{\rm He}$ fusion reactions. We then calculate the phase diagrams of the fusion cross sections with respect to the optical potential parameters and  demonstrate
 several  narrow shape resonance belts. It implies that a small deviation of  ACWFs from the exact CWFs at nuclear radius might lead to fall off the resonance regimes and therefore causes the big difference on the optical parameters as well as the cross sections.
\end{abstract}



\begin{keyword}
Light nuclear fusion \sep Coulomb wave functions \sep Optical potential parameters


\end{keyword}

\end{frontmatter}


\section{Introduction}
\label{}
The fusion cross section (FCS) is one of the fundamental physical quantities in controlled nuclear fusion research as well as nucleosynthesis of light elements in the stellar core \cite{Kikuchi:2011,Clayton:1990,Adelberger:1998,Balantekin:1998}. In the Gamow tunneling picture, nuclear fusion is generally believed to consist of three processes which are approximately independent of each other \cite{Atzeni:2003,Bosch:1992}. First, the wave packets of two nuclei collide with each other at a probability. Second, the closing nucleus tunnels through the repulsive Coulomb barrier. Third, the nuclei come into contact and fuse by nuclear potentials. At low energies where classical turning point is much larger than the nuclear radius, the FCS can be written in a phenomenological Gamow form as
a product of three terms based on the tunneling picture above \cite{Atzeni:2003,Bosch:1992}:
\begin{equation}\label{sigma}
\sigma(E)=S(E)\frac{1}{E}~{\rm exp}\left(-\frac{B_G}{\sqrt{E}}\right),
\end{equation}
where $E$ is the relative energy of two nuclei in the center-of-mass frame.
$B_G=Z_1 Z_2 e^2\sqrt{2 m}/4\hbar \epsilon_0$ is called the Gamow constant \cite{Gamow:1928}. The exponential term is the probability of tunneling through the nucleus Coulomb barrier.
$S(E)$ is the astrophysical~$S$-factor and represents mainly the nuclear part of fusion
reaction probability \cite{Adelberger:1998,Bosch:1992}. $S$-factor is a smooth slowly varying function of energy except for
the resonance zone. It is much accessible to extrapolate $S$-factor down to low energies
of astrophysical interest than the FCS \cite{Bosch:1992}.

Due to the lack of a full understanding of the nuclear potential during fusion, the FCS contains many parameters, which need to be obtained by fitting experimental data. Early in 1970s, Duane proposed an empirical FCS formula with five parameters which has been widely used in fusion research since then \cite{Duane:1972}. However, this 5-parameter fitting formula gives poor extrapolations of FCS to low energies \cite{Bosch:1992}. Bosch and Hale presented a reliable 9-parameter fitting formula based on R-matrix theory and thousands of experimental cross section data \cite{Bosch:1992}.

 These parameterized FCS formulas contain multiple parameters which do not clearly imply the information of nuclear potentials.
 Li et al. applied a simple complex spherical square-well optical potential model to describe the nuclear potentials of light nuclear fusion 
where an imaginary part of the nuclear potential represents the effect
 of absorption \cite{lixingzhong:2000}. This model only needs to fit three parameters with obvious physical significance and
 overcomes the Gamow tunneling insufficiencies: it shows that the tunneling and decay can no longer be independent in light nuclear fusion process and
need to be combined as a selective resonant tunneling \cite{lixingzhong:2000}.
This model has been widely exploited to investigate the FCSs, astrophysical~$S$-factors, and optical potential parameters associated with the  light nuclear fusions
  \cite{lixingzhong:2002,lixingzhong:2004,lixingzhong:2006,lixingzhong:2008,Singh:2019, Khan:2021}.
However, in these studies,  the approximate Coulomb wave functions (ACWFs) that only retain the leading terms of the  Coulomb wave functions (CWFs)\cite{Landau:1987}, are exploited  for the continuity conditions at the radius of nuclear potential well. Because of the sensitivity of the FCSs, astrophysical~$S$-factors as well as the  optical potential parameters  on the continuity conditions, using ACWFs should be carefully evaluated.
In recent work of \cite{lixingzhong:2012}, beyond the leading term approximation, the precise CWFs are exploited to investigate p+$^{6}{\rm Li}$ fusion and the optical potential parameters are obtained and discussed.

In this paper, based on precise CWFs, we attempt to calculate the FCSs in a complex spherical square-well nuclear potential. We calibrate  optical potential parameters for D+D, D+T, D+$^{3}{\rm He}$, p+D, p+$^{6}{\rm Li}$ and p+$^{7}{\rm Li}$ fusion reactions by benchmarking with experimental cross section data. We find that our calculated optical potential parameters are quite
different from  those of many previous results with ACWFs \cite{lixingzhong:2002,lixingzhong:2004,lixingzhong:2006,lixingzhong:2008,Singh:2019}.
For the optical model parameters of p+$^{6}{\rm Li}$, p+D and  D+D fusion reactions, the relative errors of the real part, imaginary part  and the nuclear radius of the optical potentials can reach  148{\%}, -621{\%} and -159{\%}, respectively.
We further uncover the mechanism underlying the difference between exact CWFs and ACWFs in calculating the FCS and the astrophysical $S$-factor.




\section{Theoretical framework}\label{method}

\begin{figure}[t]
\includegraphics[width=0.65\textwidth]{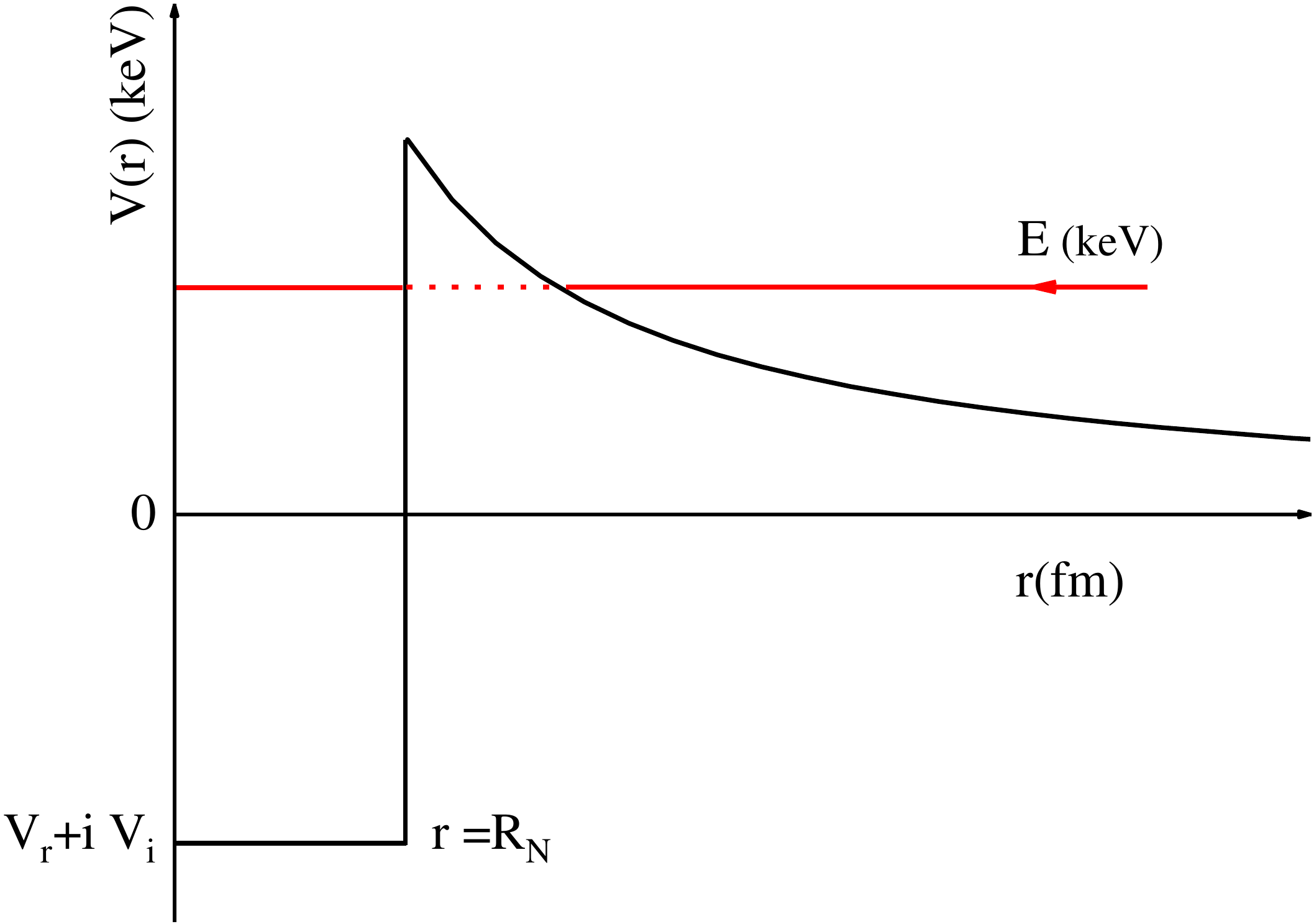}
\centering
\caption{Schematics for complex spherical square nuclear potential well and Coulomb repulsive barrier.}
\label{fig:1}
\end{figure}

According to quantum scattering theory, FCS is non-zero only for complex phase shifts. A complex potential, also known as the "optical model" \cite{Dickhoff:2018}, leads to complex phase shifts.
The real and imaginary parts of  potential represent particles scattering and absorption effects by potential, respectively. The superiority of absorptive nuclear force compared to the Coulomb repulsive leads
to an absorptive potential well for reaction in the range of nuclear force. In the simplest case, the potential is considered to be composed of a short-range complex spherical square potential well and a long-range Coulomb repulsive potential between two nuclei (Fig. \ref{fig:1}),
\begin{equation}
V(r)=\left\{
\begin{array}{rcl}
V_r+{\rm i}V_i& & {r < R_N,}\\
\frac{Z_p Z_t e^2}{4\pi \epsilon_0 r} & & {r > R_N,}\\
\end{array} \right.
\end{equation}
where $Z_pe$ and $Z_t e$ are projectile and target nuclear charges, respectively. $R_N=r_0(A_t^{1/3}+A_p^{1/3})$ is radius of nuclear potential well, and $A_t$~and $A_p$ are the mass numbers of nuclei, respectively.

According to quantum mechanics, the wave function $\psi_r(\boldsymbol{r})$ describing the relative motion of two interacting nuclei in the center-of-mass frame can be given by solving the time independent Schr\"{o}dinger equation. Here
\begin{equation}
-\frac{\hbar^2}{2 m_r}\nabla^2\psi_r(\boldsymbol{r})+(V (\boldsymbol{r})-E)\psi_r(\boldsymbol{r})=0,
\end{equation}
where $m_r=m_p m_t/(m_p+m_t)$ is reduced mass. $m_p$ and $m_t$ are mass of incident projectile and target nuclei, respectively.
It is noted that the kinetic energy of incident projectile in the laboratory system,
\begin{equation}
E_{\rm lab}=\frac{m_p+m_t}{m_t}E.
\end{equation}
In a central potential,  the wave function can be written by $\psi_r(r,\theta,\phi)=Y_{l,m}(\theta,\phi)u_l(r)/r$. The radial wave function satisfies the equations:
\begin{equation}\label{ur1}
\begin{split}
\frac{d^2u_l}{dr^2}+\frac{2m_r}{\hbar^2}\left(E-V(r)-\frac{l(l+1)\hbar^2}{2m_rr^2}\right)u_l(r)=0,\\
 u_l(r=0)=0.
\end{split}
\end{equation}

For low-energy fusion reaction, only the S-wave ($l=0$) is considered. The solution that satisfies the Eq. \eqref{ur1} can be written as \cite{Landau:1987}
\begin{equation}
u_0(r)=\left\{
\begin{aligned}
&B{\rm sin}(k_N r) ~r<R_N,\\
&D\left[F_0(kr,\eta ){\rm cot}(\delta_0)+G_0(kr,\eta)\right]~ r>R_N,\\
\end{aligned}
\right.
\end{equation}
where $B$ and $D$ are constant coefficients, $k_N=\sqrt{2 m_r (E-V_{\rm r}-{\rm i}V_{\rm i})/\hbar^2}$ is the complex nuclear wave number, $\delta_0$ is the phase shift of S-wave, $k=\sqrt{2 m_r E/\hbar^2}$ is the free particle wave number, $F_0(kr,\eta)$ and $G_0(kr,\eta)$ are regular and irregular CWFs, respectively. $\eta=1/k a_c$ is dimensionless Coulomb parameter, $a_c=4\pi\epsilon_0\hbar^2/m_r Z_p Z_t e^2$ is Coulomb unit length.
The leading terms of CWFs for small $r$ ($r\ll1,kr\ll1$) \cite{Landau:1987} are
\begin{equation}\label{CWA}
\begin{split}
&F_0(kr,\eta)\approx A k r(1+r/a_c+...),\\
&G_0(kr,\eta)\approx \frac{1}{A}\{1+2 r/a_c[{\rm ln}(2 r/a_c)+2 C-1+h(k a_c)]+...\},\\
&A^2=\frac{2\pi/k a_c}{{\rm e}^{2\pi/ka_c}-1},\\
&h(k a_c)=\frac{1}{(k a_c)^2}\sum_{j=1}^\infty\frac{1}{j(j^2+(k a_c)^{-2})}-C+{\rm ln}(k a_c),
\end{split}
\end{equation}
respectively, where $C=0.577...$ is Euler's constant.

The continuity conditions of wave function and its first derivative are satisfied simultaneously at $r=R_N$, one can obtain
\begin{equation}\label{Con_eq}
R_N k_N{\rm cot}(k_N R_N)=k R_N\frac{F_0^{'}(k R_N,\eta){\rm cot}(\delta_0)+G_0^{'}(k R_N,\eta)}{F_0(k R_N,\eta){\rm cot}(\delta_0)+G_0(k R_N,\eta)}.
\end{equation}
From Eq. \eqref{Con_eq}, phase shift $\delta_0$ can be obtained as a complex quantity and it is
convenient to assume
\begin{equation}\label{Wri}
{\rm cot}(\delta_0)\equiv W_r+{\rm i}{W_i}=\frac{k_N {\rm cot}(k_N R_N)G_0(kR_N,\eta)-G_0^{'}(kR_N,\eta)k}{k F_0^{'}(k R_N,\eta)-k_N{\rm cot}(k_N R_N)F_0(kR_N,\eta)}.
\end{equation}

It is worth pointing out that ACWFs with only retaining the leading terms (i.e., Eq. \eqref{CWA}) are exploited for continuity conditions at $r=R_N$, one can obtain \cite{lixingzhong:2000},
 \begin{equation}\label{CWAL}
\begin{split}
&R_N k_N{\rm cot}(k_N R_N)\approx \frac{R_N}{a_c}\left\{\frac{1}{\theta^2}{\rm \cot}(\delta_0)+2\left[{\rm ln}\left(\frac{2 R_N}{a_c}\right)+2C+h(k a_c)\right]\right\},\\
&{\rm cot}(\delta_0)\approx \theta^2\left\{k_N a_c {\rm cot}(k_N R_N)-2\left[{\rm ln}\left(\frac{2 R_N}{a_c}\right)+2C+h(k a_c)\right]\right\},\\
\end{split}
\end{equation}
where $\theta^{-2}=k a_c A^2$ is the Gamow penetration factor.
Then, the FCS can be put in the form,
\begin{equation}\label{sigma_0}
\sigma(E)=\frac{\pi}{k^2}(1-|e^{2i\delta_0}|^2)=\frac{\pi}{k^2}\left[\frac{-4W_i}{W_r^2+(W_i-1)^2}\right].
\end{equation}
Combining Eqs. \eqref{sigma} and \eqref{sigma_0}, the astrophysical~$S$-factor can be defined as \cite{Adelberger:1998,Bosch:1992}
\begin{equation}\label{SE}
S(E)\equiv\frac{\sigma E}{{\rm exp}\left(-B_G/\sqrt{E}\right)}=\frac{-8\pi m_r}{\hbar^2 \left(-B_G/\sqrt{E}\right)}\left[\frac{W_i}{W_r^2+(W_i-1)^2}\right].
\end{equation}

\section{Numerical results and discussion}\label{result1}
There are three optical potential parameters, $V_{\rm r}$, $V_{\rm i}$ and $r_{\rm 0}$ in our model, which are adjusted to meet the experimental benchmark FCS data. Comparing with experimental data, we calibrate optical potential parameters associated with D+D, D+T, D+$^{3}{\rm He}$, p+D, p+$^{6}{\rm Li}$ and p+$^{7}{\rm Li}$ fusion reactions using Eqs. \eqref{Wri} and \eqref{sigma_0} and results are provided in Table \ref{table1}. In our paper, the benchmark experimental data are from Ref. \cite{lixingzhong:2002} for D+T fusion reaction, Ref. \cite{lixingzhong:2008} for  D+$^{3}{\rm He}$ and D+D fusion reactions, Ref. \cite{Singh:2019} for p+D and p+$^{7}{\rm Li}$ fusion reactions and Ref. \cite{lixingzhong:2012} for p+$^{6}{\rm Li}$ fusion reaction, respectively. The nonlinear least-square method is applied to find optimal parameters. Meanwhile, for comparison, we also show the optical parameters from previous paper \cite{Singh:2019} calibrated with Eqs. \eqref{CWAL} and \eqref{sigma_0} in the Table \ref{table1}.

\begin{table}[b]
\centering
\caption{Values of optical potential parameters calibrated with CWFs for fusion reactions. Here, the
Values calibrated with ACWFs are from Ref. \cite{Singh:2019} and the relative error is defined as $\chi=\left(x_{\rm ACMFs}-x_{\rm CWFs}\right)/x_{\rm CWFs}\times 100\%$. }
\vspace{0.2cm}
\label{table1}
\renewcommand\arraystretch{1.3}
\setlength{\tabcolsep}{3.0mm}{
\begin{tabular}{ c c c c c c c c c c c }
\hline
\hline
\multicolumn{1}{c}{\multirow{2}{*}{Reaction}} &\multicolumn{3}{c}{$V_{\rm r}$(MeV)}&\multicolumn{3}{c}{$V_{\rm i}$(keV)} & \multicolumn{3}{c}{$r_{\rm 0}$(fm)}\\
\cline{2-10}
\multicolumn{1}{c}{}&CWFs&ACWFs&$\chi$&CWFs&ACWFs&$\chi$&CWFs&ACWFs&$\chi$\\
\hline
\multicolumn{1}{l}{D+D}&-60.00&-48.52&-19.13{\%}&-186.1&-263.3&41.48{\%}&1.074&2.778&-158.9{\%}\\
\multicolumn{1}{l}{D+T}&-30.00&-40.69&35.63{\%}&-49.64&-109.18&120.0{\%}&1.338&1.887&41.03{\%}\\
\multicolumn{1}{l}{D+$^{3}{\rm He}$}&-16.74&-11.86&-29.15{\%}&-43.34&-259.02&497.6{\%}&2.935&3.331&13.49{\%}\\
\multicolumn{1}{l}{p+D}&-55.0&-55.0&0&-0.00326&-0.0235&-621{\%}&1.420&1.177&-17.11{\%}\\
\multicolumn{1}{l}{p+$^{6}{\rm Li}$}&-17.05&-42.25&147.8{\%}&-2661&-7500&181.8{\%}&1.433&1.180&-17.66{\%}\\
\multicolumn{1}{l}{p+$^{7}{\rm Li}$}&-90.0&-85.0&-5.56{\%}&-300.0&-395.0&31.67{\%}&1.270&1.330&4.724{\%}\\
\hline
\end{tabular}}
\end{table}
\begin{figure}[t]
\includegraphics[width=0.75\textwidth]{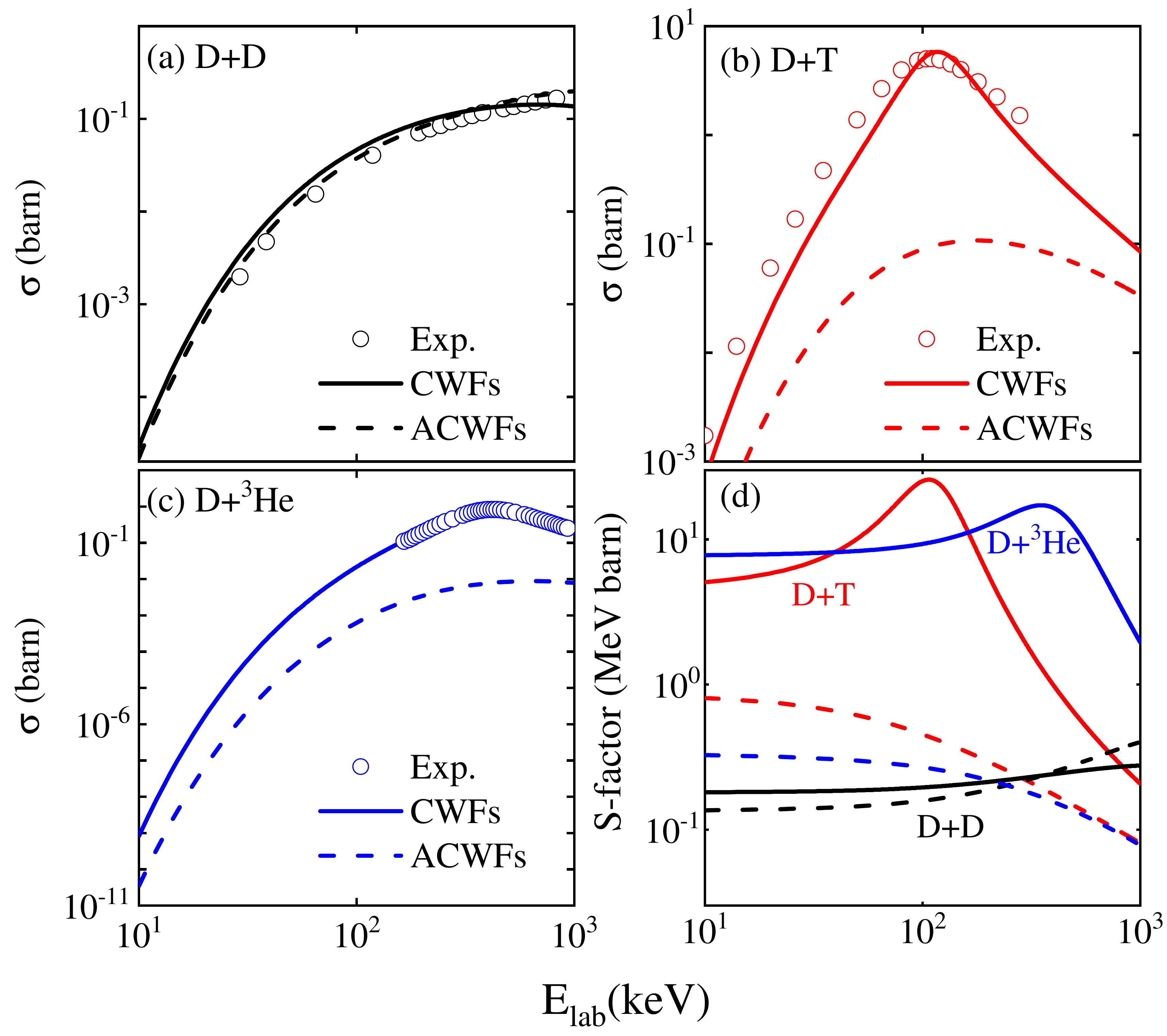}
\centering
\caption{(Color online) (a)-(c): The FCSs  for D+D, D+T and D+$^{3}{\rm He}$ fusion reactions as a function of  the energy of incident projectile in the laboratory system $E_{\rm lab}$ using our calculated optical
potential parameters; (d): Corresponding astrophysical~$S$-factors with respect to $E_{\rm lab}$: the CWFs results (solid lines) and the ACWFs results (dotted lines). The hollow circles represent the experimental data points from Ref. \cite{lixingzhong:2002, lixingzhong:2008}. }
\label{fig:2}
\end{figure}
We find that our calculated optical potential parameters with CWFs are quite  different from those using ACWFs, and the relative errors $\chi=\left(x_{\rm ACMFs}-x_{\rm CWFs}\right)/x_{\rm CWFs}\times 100\%$  between them  are shown in the Table \ref{table1}.

Furthermore, with our caculated optical potential parameters, we compare the FCSs and astrophysical S-factors (in units of MeV barn) with that formulated from ACWFs approach, and also find apparent deviation. Further
discussions on these results will be given later. For the convenience, we divide the fusion reactions into two groups for discussion in detail: the incident projectile of one group is deuteron and the other group is proton.
\subsection{ Deuteron as incident projectile}

 First, we analyze and discuss fusion reactions in which the incident projectile is deuteron.
 Compared to the previous studies using ACWFs, the real parts of our calculated optical parameters for D+D, D+T and D+$^{3}{\rm He}$ fusion reactions increase about -19\%, 36\% and 29\%, respectively. The imaginary parts of our results increase more significantly: the relative error of imaginary part for D+$^{3}{\rm He}$ fusion can reach 498\%, which implies that imaginary part is more sensitive to wave function forms for the continuity condition. More interestingly, for the radius parameter $r_0$, it is 1.074 fm for D+D reaction that is the smallest one among the three kinds of fusion reactions involved in deuteron. This is contrary to previous studies where the  radius parameter reaches minimum for D+T reaction \cite{Singh:2019}. Because the radius parameters somehow reflect the nuclear radiuses involved in fusion to some extent, considering the D nucleus is smaller than T nucleus, our results with precise CWFs seems more reasonable.

   With our calibrated optical potential parameters, the  FCSs and  astrophysical~$S$-factors as a function of the energy of incident projectile in the laboratory system are shown for D+D, D+T and D+$^{3}{\rm He}$ fusion reactions in Figs. \ref{fig:2} (a)-(d), respectively.
 One can see that the theoretical curves (solid lines) calculated with CWFs agree well with the experimental data fusion reaction.
 \begin{figure}[t]
\includegraphics[width=0.75\textwidth]{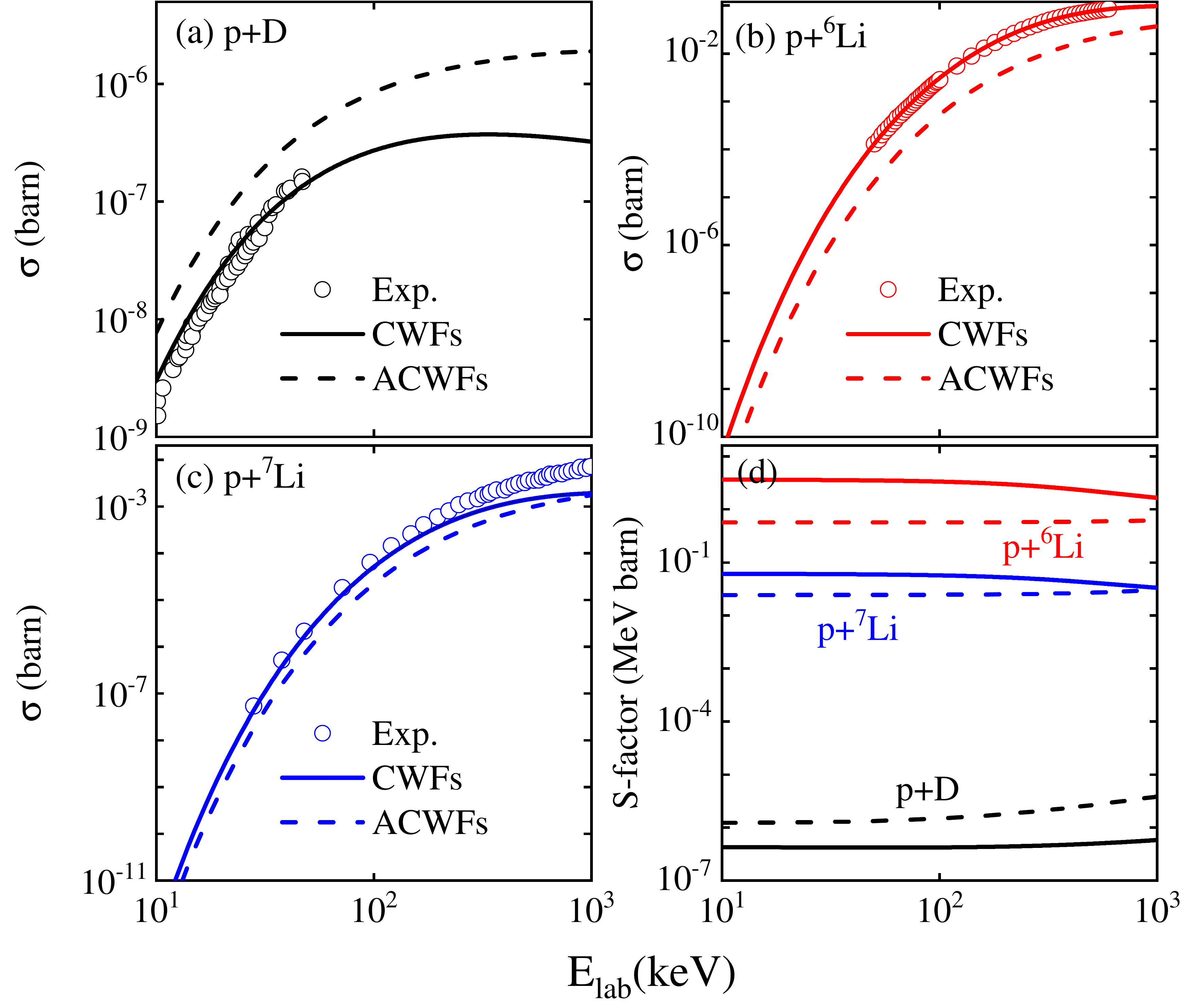}
\centering
\caption{(Color online) Same as Figs. \ref{fig:2} but for p+D, p+$^{6}{\rm Li}$ and p+$^{7}{\rm Li}$ fusion reactions and the experimental data points from Ref. \cite{Singh:2019,lixingzhong:2012}. }
\label{fig:3}
\end{figure}
 In contrast,  the curves (dotted lines) calculated with ACWFs approach deviates markedly from experimental data especially for the fusion reactions with resonance peaks such as D+T and D+$^{3}{\rm He}$. From Fig. \ref{fig:2} (d), we see that the deviations are mainly due to the discrepancies in calculating the  astrophysical~$S$-factors: in our calculations, the ~$S$-factors show apparent resonant structures for D+T and D+$^{3}{\rm He}$, while using ACWFs approach, the resonance peaks are absent, instead, two monotonously descending curves with respect to the incident energy present.

\subsection{Proton as incident projectile}

At present, we focus on results associated with p+D, p+$^{6}{\rm Li}$ and p+$^{7}{\rm Li}$ fusion reactions. 
For p+D fusion reaction, the Table. \ref{table1} presents that real part of our results is exactly equal to that of ACWFs while the imaginary part becomes smaller and the radius parameter becomes larger.
The reason for extremely small imaginary part ($V_{\rm i}=3.26$ eV) is that deuteron easily
disintegrates into p+n causing very small FCS for p+D fusion reaction and correspondingly very small $V_{\rm i}$ \cite{Singh:2019}.

Compared with optical potential parameters with ACWFs, we find that for p+$^{6}{\rm Li}$ fusion reaction, both the real part and imaginary part of our calibrated parameters are
significantly reduced: the relative errors of the real part and imaginary part are 147.8\% and 181.8\%, respectively. A recent work also obtains a set of optical potential parameters calibrated with CWFs for p+$^{6}{\rm Li}$ fusion reaction: $V_{\rm r}=-334$ MeV, $V_{\rm i}=-71.3$ MeV and $r_{\rm0}=1.36$ fm \cite{lixingzhong:2012}. Surprisingly, this set of parameters are quite different from those our calculated as well as the ACWFs. They satisfy the resonant tunneling condition near zero energy, which implies a low energy resonant tunneling in the p+$^{6}{\rm Li}$ fusion reaction \cite{lixingzhong:2012, lixingzhong:2016}. In contrast, our calculated parameters are far from the resonant tunneling condition near zero energy. Which of these two sets of parameters based on CWFs is more reasonable requires further research and discussion.
The imaginary part of potential well for  p+$^{7}{\rm Li}$ fusion reaction is significantly smaller than that of  p+$^{6}{\rm Li}$ fusion reaction, which can be explained by extremely low lifetime of $^{8}{\rm Be}$. It is very counterintuitive that the radius parameter of p+$^{7}{\rm Li}$ fusion reaction is smaller than that of p+$^{6}{\rm Li}$ in our calculated optical parameters, which is contrary to the results with ACWFs.

 We also plot the FCSs and astrophysical~$S$-factors with respect to the energy using our calculated parameters in the Figs. \ref{fig:3}. The Figs. \ref{fig:3} (a)-(c) show that the theoretical curves (solid lines) based on CWFs agree well with experimental data but for p+$^{7}{\rm Li}$ reaction which slightly deviate from experimental data at about $E_{\rm lab}=10^3$ keV. Contrary to D+T and D+$^{3}{\rm He}$ fusion reactions,  the curves using ACWFs approach (dotted lines) deviates slightly from experimental data especially for p+$^{6}{\rm Li}$ and p+$^{7}{\rm Li}$ fusion reactions, which implies that ACWFs at $r =R_N$ does not fail to some extent for these fusion reactions. For p+D fusion reaction, the deviation between the results with CWFs and ACWFs is  greater
 than that of p+$^{6}{\rm Li}$ and p+$^{7}{\rm Li}$ reactions: $\sigma=3.2\times10^{-7}$ barn for CWFs while $\sigma=1.9\times10^{-6}$ barn for ACWFs at $E_{\rm lab}=10^3$ keV. We discuss CWFs vs. ACWFs at $r =R_N$ for p+D fusion reaction in the \ref{result2}.
 \begin{figure}[t]
\includegraphics[width=0.75\textwidth]{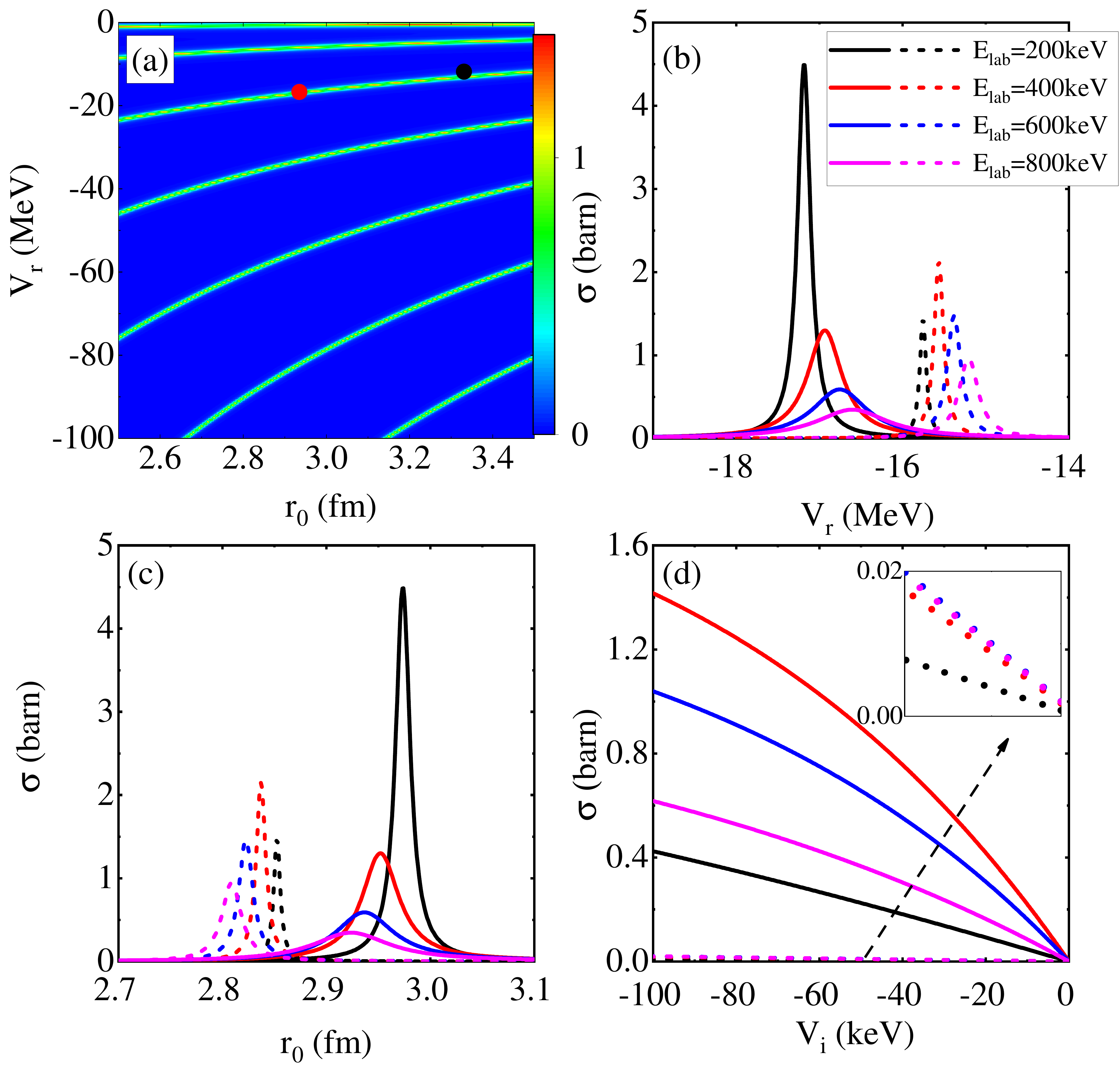}
\centering
\caption{(Color online) (a) Contour plot of FSC computed with CWFs  for D+$^{3}{\rm He}$ fusion reaction as a function of real part $V_{\rm r}$ and
radius parameter $r_{\rm 0}$ of potential well at fixed energy $E_{\rm lab}=400$ keV and imaginary part $V_{\rm i}$=-43.34 keV. The red point represents the optical parameters calibrated with CWFs while the black point represents that with ACWFs. The FSC as a function of real part $V_{\rm r}$,
radius parameter $r_{\rm 0}$ and imaginary part $V_{\rm i}$, respectively, at fixed (b) $r_{\rm 0}$=2.935fm and $V_{\rm i}$=-43.34 keV, (c) $V_{\rm r}$=-16.74 MeV and $V_{\rm i}$=-43.34 keV and (d) $V_{\rm r}$=-16.74 MeV and $r_{\rm 0}$=2.935 fm for different energies $E_{\rm lab}$=200, 400, 600 and 800 keV: the CWFs results (solid lines) and the ACWFs results (dotted lines); The subplot represents the zoom of corresponding zone in (d). }
\label{fig:4}
\end{figure}
 For the corresponding ~$S$-factors in Fig. \ref{fig:3} (d), the curves calculated with CWFs and ACWFs change slowly with energy but for slight deviations between them.

\subsection{ Shape resonance in light nuclear fusion }

In this section, we attempt to understand and explain these  results and focus on D+$^{3}{\rm He}$ fusion reaction as an example.
We study in detail the effects of three optical potential parameters $V_{\rm r}$, $V_{\rm i}$, and $r_{\rm 0}$ on FCS for D+$^{3}{\rm He}$  fusion reaction, see Figs. \ref{fig:4}.
The phase diagram of FSC computed with CWFs as a function of the real part $V_{\rm r}$ and the radius parameter $r_{\rm 0}$ is
shown in Fig. \ref{fig:4} (a) at the fixed energy $E_{\rm lab}=400$ keV and the fixed imaginary
part of potential well $V_{\rm i}$=-43.34 keV.

One can find seven separate bright belts and FCS have a larger value in the belt zone from Fig. \ref{fig:4} (a), which implies shape resonance of FCS. Meanwhile, we mark the optical potential parameters on the Fig. \ref{fig:4} (a).
The red point represents the optical potential parameters calibrated with CWFs while the black point represents that with ACWFs. The black point is exactly in shape resonance regime while the red point slightly deviates from it, which causes the big difference on optical potential parameters.

We plot FCSs with the real part and radius parameters of the potential well at fixed parameters  (b) $r_{\rm 0}$=2.935fm, $V_{\rm i}$=-43.34 keV and (c) $V_{\rm r}$=-16.74 MeV, $V_{\rm i}$=-43.34 keV respectively for different energies in the Fig. \ref{fig:4}: the solid lines are the CWFs results and the dotted
lines are ACWFs results.
 As shown in Figs. \ref{fig:4} (b)-(c), the CWFs (solid lines) and ACWFs (dotted lines) results have the same number of clear resonance peaks, but the positions and values of peak are largely different for a certain energy $E_{\rm lab}$. We can
 also see from Figs. \ref{fig:4} (b)-(c) that the resonance regions of CWFs and ACWFs hardly overlaps for different energies $E_{\rm lab}$.

The curves of FCS as a function of imaginary part of potential well at fixed parameters $V_{\rm r}$=-16.37 MeV and $r_{\rm 0}$=2.935 fm are also shown in Fig. \ref{fig:4} (d). For a fixed energy $E_{\rm lab}$, the FCS monotonically decreases with $V_{\rm i}$ and the dotted lines deviate obviously from the solid lines, which result from $V_{\rm r}$=-16.74 MeV and $r_{\rm 0}$=2.935 fm in the resonance regimes of CWFs instead of ACWFs.  As a supplement, we plot CWFs and ACWFs, and their first derivatives with respect to $r$ for D+$^{3}{\rm He}$ and p+D fusion reactions in the \ref{result2}, which also imply that the ACWFs are invalid at $r=R_N$.
 In short, we can conclude that a small deviation of ACWFs from the exact CWFs at nuclear radius might lead to fall out the resonance regimes and therefore causes big difference on the optical parameters, FCSs and corresponding~$S$-factors.  For other light fusion reactions, there are also similar properties.

\section{Conclusions}

The optical potential parameters, which  not only apply  to  nuclear science and industry, but also might carry some useful information on nuclear structures, therefore call for careful investigations. In the present work, based on precise CWFs, we calibrate optical potential parameters $V_{\rm r}$, $V_{\rm i}$, and $r_{\rm 0}$ for D+D, D+T, D+$^{3}{\rm He}$, p+D, p+$^{6}{\rm Li}$ and p+$^{7}{\rm Li}$ fusion reactions.
 We find that our calculated optical potential parameters are quite
different from  those of many previous results \cite{lixingzhong:2002,lixingzhong:2004,lixingzhong:2006,lixingzhong:2008,Singh:2019}, in which ACWFs with only retaining the leading terms are exploited for the continuity conditions at the radius of nuclear potential well. The phase diagrams of FCSs with respect to the optical potential parameters demonstrate several separate  narrow shape resonance belts, indicating that a small deviation from the exact CWFs might lead to fall off the resonance regimes and therefore cause the big difference on the optical parameters. Physical implications of the optical potential parameters are also discussed.

The FCS in our paper only contains the contribution of S-wave ($l=0$) and effect of the centrifugal barrier for $l>0$ should be further studies.
The complex spherical square-well is simplest optical potential, which have three optical
parameters. A Woods-Saxon nuclear potential with rigid core containing five parameters can be investigated using the CWFs \cite{Koohrokhi:2016}. Furthermore, the statistical factor of FCS dependent on spin can be studied based on CWFs for light nuclear fusion reactions \cite{Koohrokhi:2016}.

\section*{CRediT authorship contribution statement}

\textbf{Binbing Wu}: Investigation, Methodology, Software, Writing-original draft. \textbf{Duan Hao}:  Writing-review $\&$ editing.
\textbf{Jie Liu}: Resources, Supervision,  Writing-review $\&$ editing.

\section*{Declaration of competing interest}
The authors declare that they have no known competing financial interests or personal relationships that could have appeared to influence the work reported in this paper.
\section*{Acknowledgments}
We thank Dr. Wenjuan Lv and Dr. Yameng Li  for their critical reading of the paper. This work was supported by funding from NSFC No.\ 11775030 and NSAF No. U\ 1930403.


\appendix
\section{CWFs versus ACWFs}\label{result2}
In this appendix, we discuss in detail wave functions and their first derivatives with respect to $r$  for CWFs and ACWFs. The regular CWF $F_0$ can be written as \cite{Abramowitz:1972}
\begin{figure}[!b]
\includegraphics[width=0.65\textwidth]{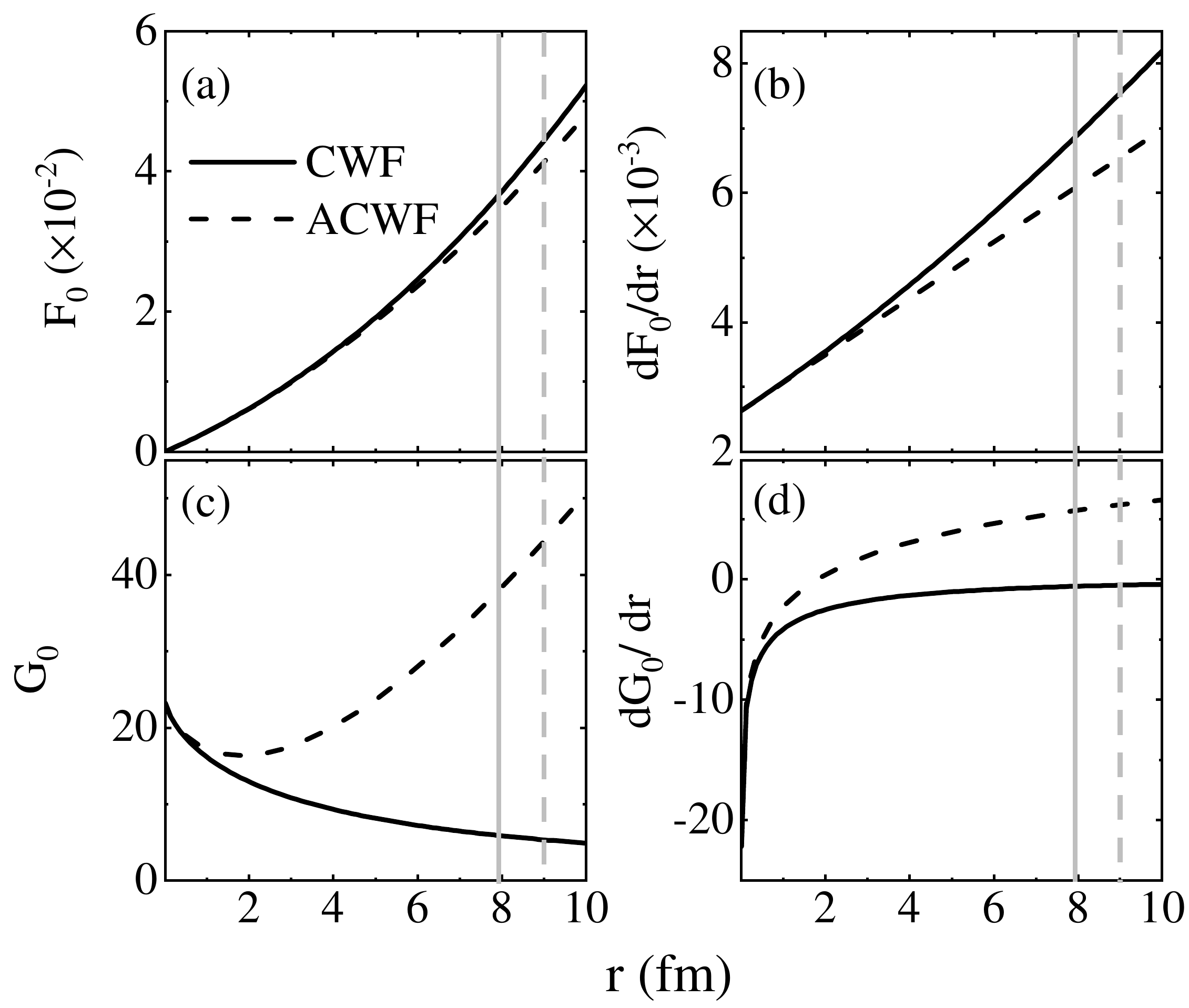}
\centering
\caption{(Color online) (a)-(d): $F_0$ and $G_0$ and their first derivative as a function
of radius at a fixed energy $E_{\rm lab}$=110 keV for D+$^{3}{\rm He}$ fusion reaction; The vertical grey lines represent $r=R_N$: the CWFs results (solid lines) and the ACWFs results (dotted lines).  }
\label{fig:5}
\end{figure}
\begin{equation}
F_0(kr,\eta)={\rm CF}_0(\eta){\rm M}\left(i\eta, \frac{1}{2}, 2ikr\right),
\end{equation}
where $
{\rm CF}_0(\eta)=-\frac{1}{2} i e^{\frac{-\pi\eta}{2}}\sqrt{\pi \eta {\rm csch}(\pi \eta)}$, ${\rm M}$ and csch represent Whittaker M and the hyperbolic cosecant functions, respectively.
The irregular CWF $G_0$ can be represented by \cite{Abramowitz:1972}
\begin{equation}
G_0(kr,\eta)=\frac{1}{2}\left({\rm CHP}_0(\eta,kr)+{\rm CHM}_0(\eta,kr)\right),
\end{equation}
where
\begin{equation}
\begin{split}
{\rm CHP}_0(\eta,kr)&={\rm W}\left(-i\eta, \frac{1}{2}, -2ikr\right)e^{i{\rm CP}_0(\eta)+\frac{1}{2}\pi\eta},\\
{\rm CHM}_0(\eta,kr)&={\rm W}\left( i\eta, \frac{1}{2}, 2ikr\right)e^{-i{\rm CP}_0(\eta)+\frac{1}{2}\pi\eta},\\
{\rm CP}_0(\eta)&=\frac{{\rm ln}[\Gamma(1+i\eta)/\Gamma(1-i\eta)]}{2i}. \\
\end{split}
\end{equation}
${\rm W}$ and $ \Gamma$ are Whittaker W and
Gamma functions, respectively.
The first derivatives with respect to $r$ for CWFs can be calculated by mean of \cite{Abramowitz:1972}
\begin{equation}
\begin{split}
\frac{\partial{\rm M}(a, b, x)}{x}&=\left(\frac{1}{2}-\frac{a}{x}\right){\rm M}(a, b, x)+\frac{\left(a+b+\frac{1}{2}\right){\rm M}(a+1, b, x)}{x},\\
\frac{\partial{\rm W}(a, b, x)}{x}&=\frac{1}{2x}\left[(x-2a){\rm W}(a, b, x)-2{\rm W}(a+1, b, x)\right].\\
\end{split}
\end{equation}
It is noted that the W and M are built-in functions in many math softwares such as matlab and CWFs and their first derivatives can be easily calculated by the formulas in the appendix.

The Fig. \ref{fig:5} presents that the CWFs and ACWFs and their first derivatives as a function of radius at a fixed energy $E_{\rm lab}$=110 keV for D+$^{3}{\rm He}$ fusion reaction. As shown in Figs. \ref{fig:5} (a)-(b), ACWF starts to deviate from CWF for
$F_0$ about $r=4$ fm and its derivative about $r=2$ fm. The difference between the ACWF and CWF for both $F_0$ and their first derivative become larger and larger as radius increases. Figs. \ref{fig:5} (c)-(d) show that the ACWF and their first derivatives of $G_0$ are valid near $r=0$ fm and largely deviate the CWF about $r=1$ fm. In our calculated optical parameters, the radius of nuclear potential well $R_N=r_0(3^{1/3}+2^{1/3})$= 7.93 fm (9.00 fm for ACWFs) for D+$^{3}{\rm He}$ fusion reaction implies that ACWFs fail for the continuity conditions at $r=R_N$.
\begin{figure}[!t]
\includegraphics[width=0.65\textwidth]{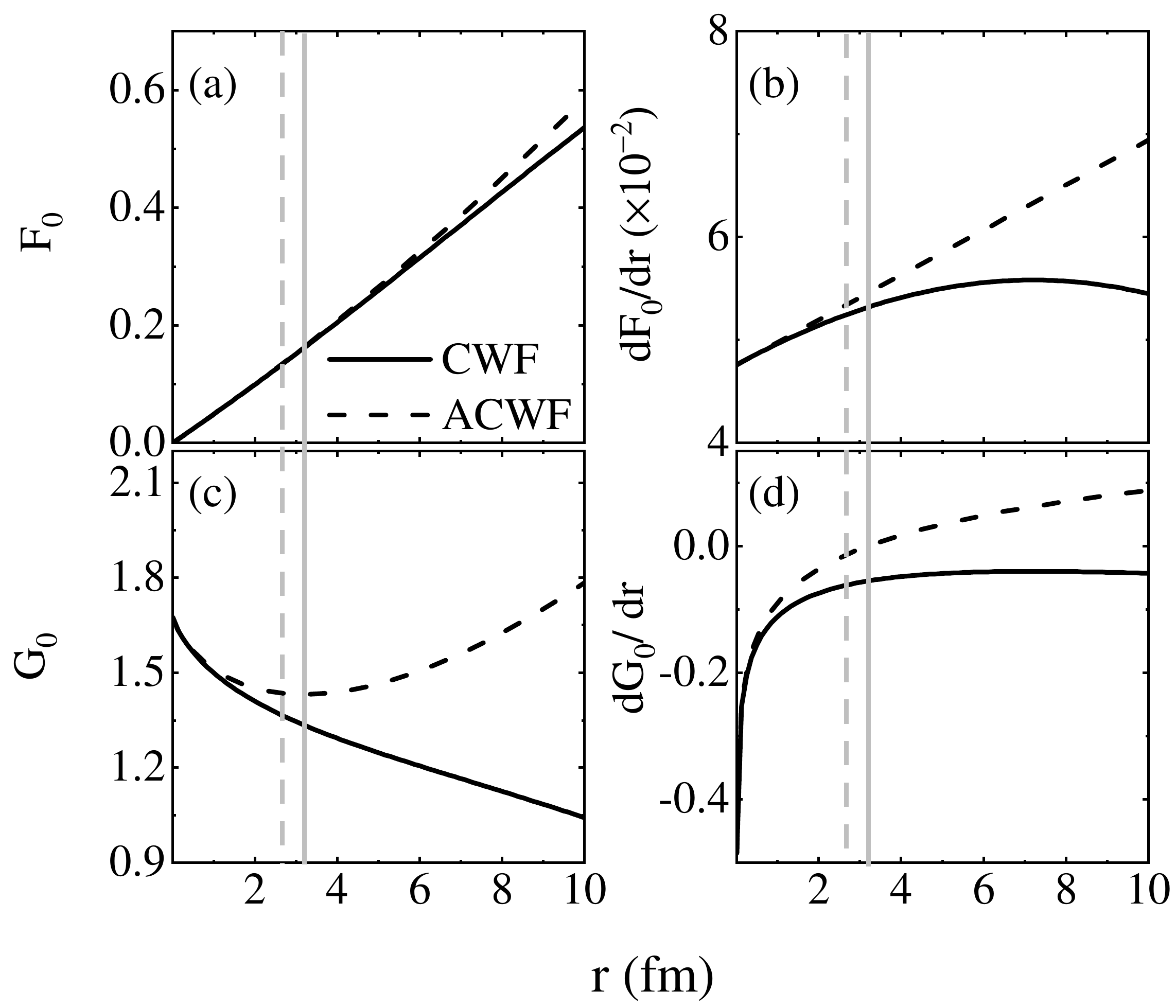}
\centering
\caption{(Color online) (a)-(d): Same as Figs. \ref{fig:5}(a)-(d) but for p+D fusion reaction and $E_{\rm lab}$=300 keV. }
\label{fig:6}
\end{figure}

We also show the case of p+D fusion reaction at $E_{\rm lab}$=300 keV in the Fig. \ref{fig:6}. The radius of nuclear potential well $R_N=r_0(1^{1/3}+2^{1/3})=3.21$ fm (2.66 fm for ACWFs) in our calculated optical parameters for p+$^{6}{\rm Li}$ fusion reaction also suggest that ACWFs are invalid at $r=R_N$ to some extent.



\end{document}